# Band Gap Opening and Optical Absorption Enhancement in Graphene using ZnO Nanoclusters

M. M. Monshi, S. M. Aghaei, and I. Calizo

*Abstract*— Electronic, optical and transport properties of the graphene/ZnO heterostructure have been explored using first-principles density functional theory. The results show that $Zn_{12}O_{12}$ can open a band gap of 14.5 meV in graphene, increase its optical absorption by 1.67 times covering the visible spectrum which extends to the infra-red (IR) range, and exhibits a slight non-linear *I-V* characteristic depending on the applied bias. These findings envisage that a graphene/$Zn_{12}O_{12}$ heterostructure can be appropriate for energy harvesting, photodetection, and photochemical devices.

## I. INTRODUCTION

Graphene as a 2D carbon allotrope has drawn tremendous attention worldwide because of its unique properties for electronics, spintronics, and surface sciences [1-2]. It is a promising material for high-performance nanoelectronics due to its high carrier concentration, mobility, and stability [3-4]. However, widespread adoption of graphene for electronic devices in particular still faces challenges because it lacks an energy band gap as a result of band degeneracy at the Dirac point. Breaking translational symmetry or sublattice symmetry could be an option to introduce a band gap in graphene. From density functional studies, it is apparent that a band gap can be opened in graphene in a variety of ways including chemical edge-functionalization [5, 6] and quantum confinement with graphene nanoribbons [5-9], nanomesh [10], periodic nanoholes [7,11], silicon doping [12], oxygen doping, strain, vacancy induced, [13] etc. Experimentally it has been proven that a band gap ranging from 2.5 meV to 450 meV can be opened in graphene with the application of strain or electric-field [14], decoration with Si-rich two-dimensional islands [15] and the introduction of defects [11], formation of perforated graphene using black copolymer (BCP) lithography [16], and adsorption of patterned hydrogen [17] and water molecules [18]. Most of these methods require precision tools or complex lithographic process to achieve band gaps that in some instances are negligible. Pala *et al.* investigated a novel process to open band gap by decorating graphene using different nanoparticles where special attention was given to ZnO nanoseeds [19].

ZnO has a wide band gap of 3.37 eV and large exciton binding energy of ~60 meV (2.4 times of the room-temperature thermal energy) [20]. Hence, the incorporation of ZnO with graphene could be a promising heterostructure for electronic and optoelectronic devices such as solar cells, field emission, displays, sensors, light emitting, and detection devices in UV-visible spectral range. It could also be used in the photocatalytic degradation of organic pollutants under UV or visible light irradiation [21, 22]. ZnO nanoparticles also possess thermal and mechanical stability and require a rapid and inexpensive synthesis technique using available cost-effective standard semiconductor device fabrication technologies [20].

The experimental absorption energies of small ZnO nanoparticles are found to be very close to the optical excitation energy of $Zn_{12}O_{12}$ cluster calculated in the framework of time-dependent density functional theory (DFT) [23]. This observation leads the authors of Ref. [23] to conjecture that the surface structure of small ZnO nanoparticles is very similar to that of a $Zn_{12}O_{12}$ nanocluster because the excitation energies of former (~3.83 eV) are in good agreement with that of later (3.83 eV). Therefore, we have study the $Zn_{12}O_{12}$ nanocluster on graphene.

In the present work, first-principles calculations are employed to investigate the effects of $Zn_{12}O_{12}$ on the electronic, transport, and optical properties of graphene. We found that the current-voltage response of the graphene/$Zn_{12}O_{12}$ heterostructure is slightly non-linear due to the occurrence of a small band gap in graphene. In addition, the optical absorption co-efficient of the heterostructure reflects enhanced optical properties.

## II. COMPUTATIONAL METHODOLOGY

Calculations are performed using first-principle methods based on DFT implemented in Atomistix ToolKit (ATK) package [24]. The exchange-correlation functional is approximated by the Generalized Gradient Approximation of Perdew-Burke-Ernzerhof (GGA-PBE) with a double-ζ polarized basis set [25-27]. First, the graphene/$Zn_{12}O_{12}$ heterostructure, consisting of 5×5 supercell of graphene and a $Zn_{12}O_{12}$ nanocluster (as shown in Fig. 1(a) and (b)), has been

This work was supported in part by Florida Education Fund's McKnight Junior Faculty Fellowship.
The authors are with the department of Electrical and Computer Engineering, Florida International University, Miami, FL, USA (e-mail: mmons021@fiu.edu)

fully relaxed (lattice constant and atomic positions) until the maximum atomic force and stress on each atom are less than 0.01 eV/Å, 0.05 eV/Å$^3$, respectively. A grid mesh cut-off energy of 75 Ha and 1×3×3 k-points for the integration of the first Brillouin zone are chosen. A vacuum separation of 15 Å perpendicular to the structure is considered to form a free-standing ultrathin film to suppress adjacent images interactions. To describe long-range van der Waals (vdW) interactions, the Grimme vdW correction (DFT-D2) is considered [28]. The k-points are increased to 1×101×101 for calculations of electronic, transport, optical properties.

The electron transport calculations are performed using GGA/PBE combined with non-equilibrium green's function (NEGF). A fast Fourier transform (FFT) solver is used to solve Poisson equation.

The schematic of the two electrode system for the graphene/Zn$_{12}$O$_{12}$ heterostructure is presented in Fig. 1(c). The electrical current through the device under non-equilibrium conditions, *i.e.* for a finite bias voltage ($V_{Bias}$), can be calculated using the Landauer formula as follows [29]

$$I(V_{Bias}) = \frac{2e}{h}\int_{-\infty}^{+\infty} T(\varepsilon, V_{Bias})[f(\varepsilon-\mu_L) - f(\varepsilon-\mu_R)]d\varepsilon \quad (1)$$

Where $f$ is the Fermi-Dirac distribution function, $\mu_{L,R}=E_F \pm eV_{Bias}/2$ represents the chemical potentials of the left and right electrodes, and $T(\varepsilon, V_{Bias})$ is the energy and voltage dependent transmission function. The transport setup consists of two graphene electrodes of 2 Å with the graphene/ Zn$_{12}$O$_{12}$ structure as central region. The supercell was oriented in such a way that the transport direction connecting L and R was aligned parallel to the armchair direction within the structure, as shown in Fig. 1(c).The optical absorption spectrum is calculated using the Kubo-Greenwood formula [30]. The susceptibility tensor in this method given as follows

$$\chi_{ij}(\omega) = -\frac{e^2\hbar^4}{m^2\varepsilon_0 V \omega^2}\sum_{nm}\frac{f(E_m)-f(E_n)}{E_{nm}-\hbar\omega-i\hbar\Gamma}\times \pi^i_{nm}\pi^j_{mn} \quad (2)$$

Where, $\pi^i_{nm}$ means $i$-component of the dipole matrix element between $n$ and $m$ states, $\Gamma$ is the broadening, $V$ is the volume, $f$ is the Fermi function and $E_m$ ($E_n$) corresponds to eigenvalues of $m(n)$ state. The dielectric constant ($\varepsilon_r$) can be written in terms of the susceptibility tensor $\chi(\omega)$, using linear response theory as follows

$$\varepsilon_r(\omega) = 1 + \chi(\omega) \quad (3)$$

The refractive index, $n$, is related to the complex dielectric constant through

$$n + i\kappa = \sqrt{\varepsilon_r} \quad (4)$$

Here, $\kappa$ is the extinction coefficient. Finally, the optical absorption coefficient is related to the extinction coefficient through [31]

$$\alpha_a = 2\frac{\omega}{C}\kappa \quad (5)$$

For optical properties calculation, Hartwigsen, Goedecker, Hutter (HGH) pseudopotential is used for exchange-correlation functional with a good description of virtual states far above the Fermi level.

### III. RESULTS AND DISCUSSION

The shortest separation distance between an atom of the adsorbed Zn$_{12}$O$_{12}$ nanocluster and the closest C atom of the graphene monolayer in the graphene/Zn$_{12}$O$_{12}$ complex is 2.86 Å which is in agreement with the previous result [32]. The bond length between Zn and O is found to be in the range of 1.86 Å-1.97 Å which concurs with reported data [33]. From the relaxed geometries of the nano-complexes, it is found that there is no significant structural change of Zn$_{12}$O$_{12}$ nanocluster with respect to their free counterparts whereas graphene undergoes a small local structural distortion.

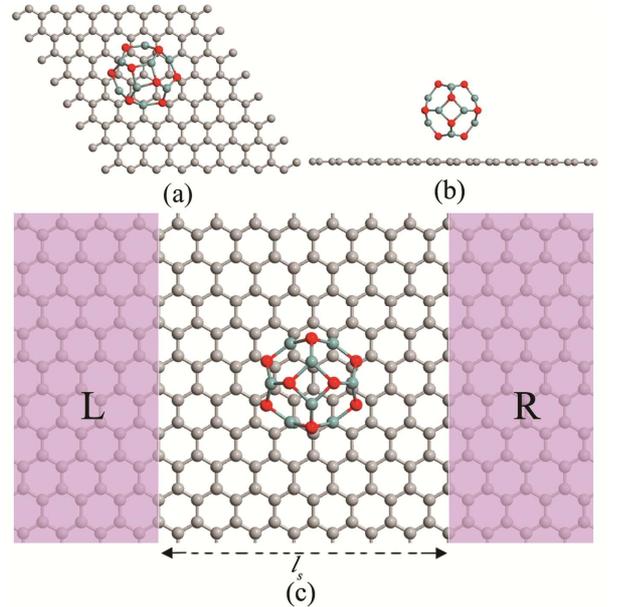

Fig. 1. Graphene/ Zn$_{12}$O$_{12}$ atomic structure (a) top view, (b) side view where gray is carbon, cyan is zinc and the red ball is oxygen, and the (c) two electrode system of graphene/ Zn$_{12}$O$_{12}$ heterostructure

The interaction between the nanocluster and the graphene layer is expected to be small. However, it was discovered that there is an effective charge transfer between the adsorbed metal/metal oxide nanopartciles and graphene [30]. Our Mulliken population analysis suggests that a net negative charge of 0.104 $e$ is transferred from the graphene layer to the nanocluster due to the differences between the electron affinities of the nanocluster and that of graphene.

Bandstructure, as well as the density of states (DOS) of the Zn$_{12}$O$_{12}$ decorated graphene complex, is considered to visualize the impact on electronic property. In neutral graphene,a zero band gap semiconductor, the lowest conduction band and the upper valence band are degenerate at Fermi level ($E_F$). The graphene band structure and corresponding DOS are significantly affected by the presence of $Zn_{12}O_{12}$ nanoclusters. As can be seen in Fig. 2, the upper valence band and lower conduction shift apart at $E_F$ due to the charge transfer. From the fully relaxed complex structure, it is clear that the graphene-nanocluster interaction region is slightly deformed which could be the reason for A-B sublattice symmetry breaking at least locally and consequently opening the band gap through a charge transfer process near the Dirac point. . Thus, 14.5 meV band gap is found near the K-point as can be seen in the Fig. 2 insets. This band gap is comparable to the experimental finding in the Ref. [19] (7.36 meV) where ZnO nanoparticles are randomly distrubuted to tune the band gap by varying the particles size and densities. Moreover, the band gap is close to the previous theoretical finding (25 meV) [32] but did not match exactly. The reason might be vdW interaction which was not included in the previous calculations.

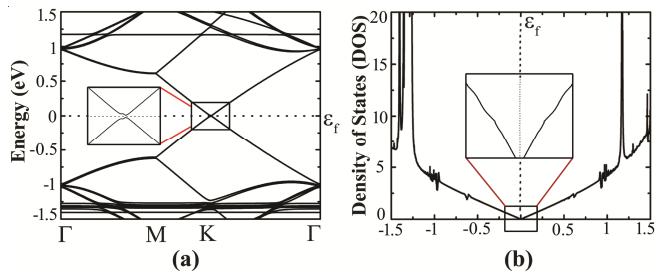

Fig. 2. (a) Bandstructure and (b) density of states of graphene/ $Zn_{12}O_{12}$

Optical abosprtion coefficient study results in Fig. 3(a) show that the absorbance increased by 1.67 times while maintaining the same characteristics as pure graphene covering the visible and infra-red (IR) spectrum. This execellent photo responsiveness of the graphene/$Zn_{12}O_{12}$ heterostructure could make it a promising candidate for solar energy and photo detecting devices. Because the $Zn_{12}O_{12}$ nanocluster opens a band gap in graphene, it is logical to claim that the structural complex could generate significant current once the device is fabricated and the bias is applied in a simple two electrode device. The I-V characteristics of the graphene/$Zn_{12}O_{12}$ shows ambipolar behavior where forward and reverse bias produces similar current in the device asseen in Fig. 3(b). Instead of linear behavior like pristine graphene [2], a slight nonlinear I-V characteristic is apparent for the energy gap because of $Zn_{12}O_{12}$ nanocluster induced symmetry breaking in graphene [34].

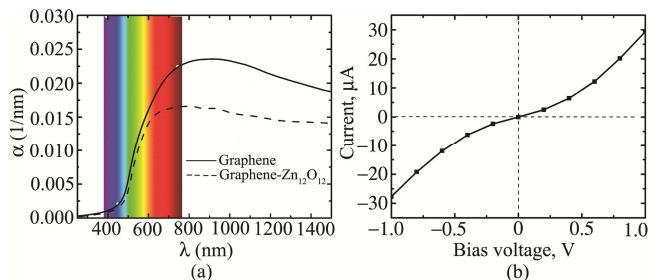

Fig. 3. (a) Optical absorption coefficient and (b) I-V characteristics of graphene/ $Zn_{12}O_{12}$ two electrode system.

## IV. CONCLUSION

First-principles calculation envisaged the influence on electronic, transport and optical properties induced by a $Zn_{12}O_{12}$ nanocluster on graphene. The results show that this nanocluster opens a sizable band gap as a result of charge transfer. The optical absorption coefficient increases within the visible range and beyond while maintaining the similarity in graphene's characteristics making the heterostructure useful for potential solar energy and photo detecting devices. The *I-V* characteristics of the graphene/$Zn_{12}O_{12}$ heterostructure exhibit a slight nonlinear behavior because of symmetry breaking. Increasing nanocluster density on graphene could provide a practical way to open sizable band gap and tune it for graphene.